\documentclass[dvips]{article}

\usepackage[dvips]{graphicx}

\begin{document}


\newcommand{\nn}{\nonumber}
\newcommand{\bra}{\langle}
\newcommand{\ket}{\rangle}
\newcommand{\DA}{\centerline{{$\Downarrow$}}}
\newcommand{\ds}{\displaystyle}



\title{%
Neutrino Oscillation
 and
CP Violation
}

\author{%
Joe Sato\\
  Research Center for Higher Education,
 Kyushu University,\\
Ropponmatsu, Chuo-ku, Fukuoka, 810-8560, Japan.
}

\date{}
\maketitle

\begin{abstract}
We reconsider the meaning of observing CP violation
in neutrino oscillation.
\end{abstract}

\section{Introduction}

Many experiments and observations have shown evidences for neutrino
oscillation one after another.  The solar neutrino deficit has long
been observed\cite{Ga1,Ga2,Kam,Cl,SolSK}.  The atmospheric neutrino
anomaly has been found\cite{AtmKam,IMB,SOUDAN2,MACRO} and recently
almost confirmed by SuperKamiokande\cite{AtmSK}.  All of them can be
understood by neutrino oscillation and hence indicates that neutrinos
are massive and there is a mixing in lepton
sector\cite{FukugitaYanagida}. Relevant parameters will be
determined more precisely in near future\cite{K2K}.

Thus  completely unknown parameters for
the lepton sector will be 

\begin{center}
\begin{tabular}{ll}
$U_{e3}$ &: Last Mixing\\
$\sin\delta$ &: CP Violation\\
sign  of $\delta m_{\rm atm}^2$&
\end{tabular}
\end{center}

\noindent
in near future.

Then how we determine them is a big problem and it is the main topic 
of the conference. In this article we will pay attention to CP violating
phase.

What energy range is suitable for observing CP violation?
 Since CP-violation effect arise as three(or more)-generation
phenomena\cite{KM}, we should make an experiment
with ``not too high'' and ``not too low'' energy to see ``3-generation''.
In an oscillation experiment, there are two energy scales,
\begin{eqnarray}
E\ \sim\ 
\left\{
\begin{array}{l}
\ \delta m^2_{31} L \\
\ \delta m^2_{21} L
\end{array}
\right. .
\label{energyRange}
\end{eqnarray}
Then the above energy range is expected to be suitable for
a neutrino oscillation experiment to see CP violation in lepton
sector\cite{Joe}.

Indeed
in high energy the first two lightest states seem ``degenerate''.
\begin{eqnarray}
 \Longleftrightarrow \delta m^2_{21} \sim 0
\nonumber
\end{eqnarray}
and hence the oscillation term for CP violation becomes 0:
\begin{eqnarray}
&&\sin\Delta_{21}+\sin\Delta_{32}+\sin\Delta_{13}
\nonumber
\\
&\sim& \sin\Delta_{31}+\sin\Delta_{13} \Longrightarrow 0.
\nonumber
\end{eqnarray}

In low energy the heaviest (two) state(s) ``decouple(s)''.
\begin{eqnarray}
 \Longleftrightarrow \Delta m^{2'} s \sim \infty
\nonumber
\end{eqnarray}
and therefore the oscillation term is averaged away within
finite resolution for neutrino energy:
\begin{eqnarray}
\begin{array}{cccccc}
&\sin\Delta_{21}&+&\sin\Delta_{32}&+&\sin\Delta_{13}.\\
\hbox{ oscillating out}&
\downarrow&&\downarrow&&\downarrow\\
&0&&0&&0
\end{array}
\nonumber
\end{eqnarray}

From this consideration,
\begin{eqnarray}
\delta m^2_{21} L \leq E \leq  \delta m^2_{31} L
\label{bestEnergy} 
\end{eqnarray}
will be the best energy range.

Moreover, to avoid the uncertainty due to matter effect\cite{MSW},
lower energy and shorter baseline length
are better.

Experimentally there are two energy region for neutrino 
experiment\cite{Konaka}.
One is $E_\nu \sim$ 0.1-1  GeV and
the other $E_\nu > 5$ GeV which is considered extensively
in the context of neutrino factory.

\section{
Oscillation probability {$
P(\nu_\alpha\rightarrow\nu_\beta)$}\\
for  $E \sim$ 0.1-1 GeV and  $L \sim$ O(100) Km
}

In this subsection we will consider the neutrino oscillation experiment
with $E_\nu \sim$ 0.1-1 GeV. For this energy region the suitable
baseline length $L$ to see CP violation is determined to be
on the order of 100 km by the ``3 generation
condition'' (\ref{energyRange}).

For this setting, the transition probability is calculated to be
\begin{eqnarray}
 & & P ( \nu_\mu \rightarrow \nu_e ) \nonumber \\
 &=& 4 |U_{e3}U_{\mu 3}|^2 \sin^2 \frac{\Delta_{31}}{2}
\nonumber\\
 &+& 4 {\rm Re}(U_{e3}^*U_{\mu 3} U_{e2} U_{\mu 2}^*) 
      \left( \frac{\delta m_{21}^2}{\delta m_{31}^2} \right)
      \Delta_{31} \sin \Delta_{31}
\nonumber\\
{\rm  CPV !!}
 &-& 4 {\rm Im}(U_{e3}^*U_{\mu 3} U_{e2} U_{\mu 2}^*) 
      \left( \frac{\delta m_{21}^2}{\delta m_{31}^2} \right)
      \Delta_{31} \sin^2 \frac{\Delta_{31}}{2}
\nonumber\\
&-&  4  {\rm Re}(U_{e2}^*U_{\mu 2} U_{e1} U_{\mu 1}^*)
  \left(\frac{\delta m_{21}^2}{\delta m_{31}^2}\right)^2
  \left(\frac{\Delta _{31}^2}{2}\right)^2
\nonumber\\
&\equiv&
A \sin^2 \frac{\Delta_{31}}{2}
\nonumber\\
 &+& \frac{B}{2} \Delta_{31} \sin \Delta_{31}
\nonumber\\
 &+& C    \Delta_{31} \sin^2 \frac{\Delta_{31}}{2}
\label{transitionP}\\
&+&  D  \left(\frac{\Delta _{31}^2}{2}\right)^2
\nonumber
\end{eqnarray}
up to the leading(second) order of small values,
\begin{center}
 $U_{e3}$,
$      \frac{\delta m_{21}^2}{\delta m_{31}^2} $ and
$      \frac{a}{\delta m_{31}^2} ,$
\end{center}
here $a$ denotes the matter effect.

There are two comments here: 
1) What we can observe are not mixing
angles of a certain parameterization but values of certain combination of
couplings which are $A-D$ in this case. Without paying attention to
this fact, we will not understand correctly the uncertainties
on the mixing parameters due to uncertainties of an experiment.
2) For this setting the matter effect, which is the serious
obstacle for detecting CP violation, gives only a subleading
effect, and in this sense the neutrino energy and the baseline
length assumed here seems very preferable. To observe
asymmetry in the transition probability means directly
the fact that there is a CP violation in lepton sector.

Current bounds on coefficients\cite{Fogli}:
\begin{eqnarray}
A &\leq& 0.05\nonumber\\
B&\leq& 0.006\nonumber\\
C&\leq& 0.006\nonumber\\
D&\leq& 0.001\nonumber
\end{eqnarray}
with  $U_{e3}\leq 0.15$ and 
$\frac{\delta m_{21}^2}{\delta m_{31}^2} <3\times 10^{-2}.$
Due to the different energy dependence, these
four terms can contribute the oscillation probability equivalently!!

In eq.(\ref{transitionP}),
base functions,
\begin{eqnarray} 
\sin^2\frac{\Delta_{31}}{2},\  \Delta_{31}\sin\Delta_{31},
\ \Delta_{31}\sin^2\frac{\Delta_{31}}{2},\  \Delta_{31}^2
\nonumber
\end{eqnarray}
are independent! 
Since
\begin{eqnarray}
 \left\{
\begin{array}{ccl}
 L&= &300{\rm km} \\
\delta m^2_{31}&=&3\times 10^{-3}
\end{array}
\right.
\Longleftrightarrow
\frac{\Delta_{31}}{2}=
\left\{
\begin{array}{lcl}
 \frac{1}{2}\pi&{\rm at} &E\sim 700 {\rm MeV} \\
 \frac{3}{2}\pi&{\rm at} &E\sim 250 {\rm MeV}
\end{array}
\right.
\nonumber
\end{eqnarray}
in the energy region considered here their
behaviors are completely different from each other and
hence it is expected that the coefficients $A-D$ are
determined rather well.

\begin{figure}[h]
\unitlength 1cm
\begin{picture}(7,5)(0,0)
\put(0,4){$\displaystyle  \sin^2 \frac{\Delta_{31}}{2}$}
\put(6,-0.2){$\displaystyle  \Delta_{31}$}
  \includegraphics[width=6cm,height=4cm,clip]{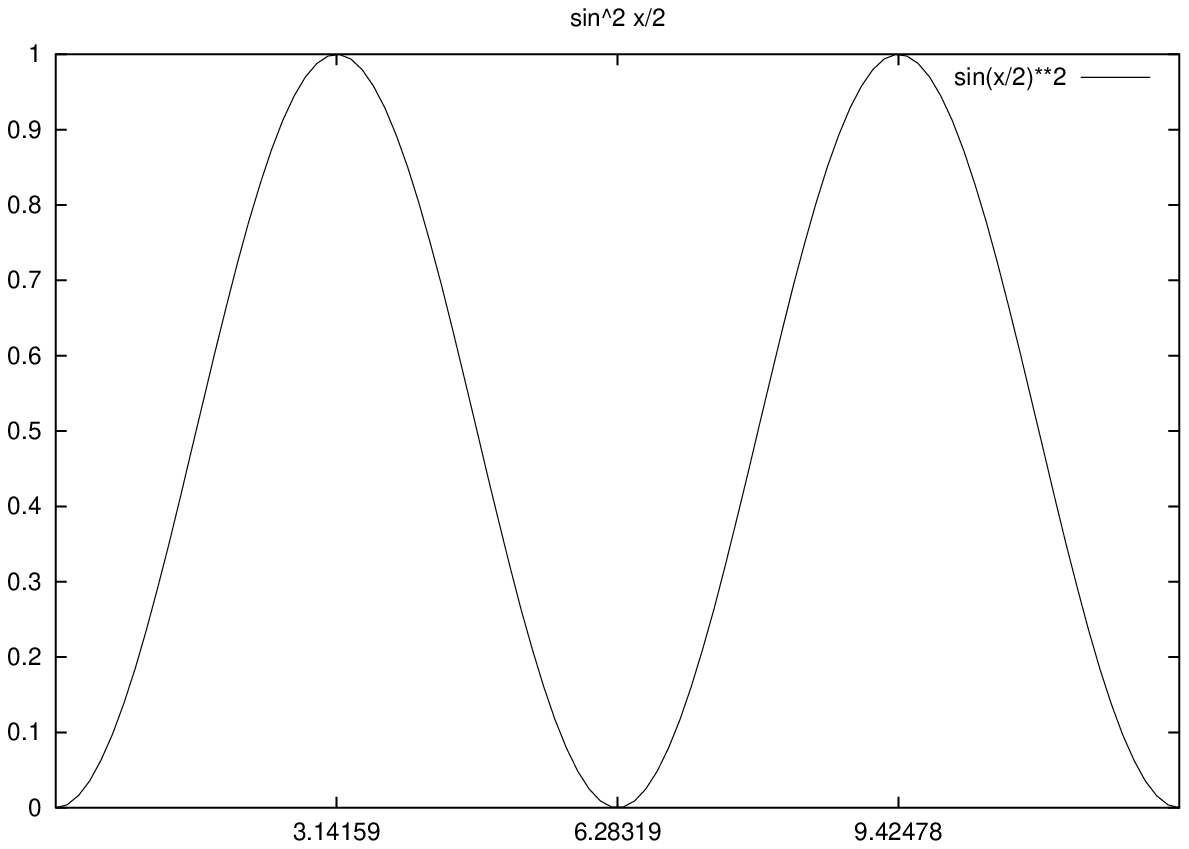}
\end{picture}
\begin{picture}(7,5)(0,0)
\put(0,4){$\displaystyle  \Delta_{31}\sin^2\frac{\Delta_{31}}{2}$}
\put(6,-0.2){$\displaystyle  \Delta_{31}$}
  \includegraphics[width=6cm,height=4cm,clip]{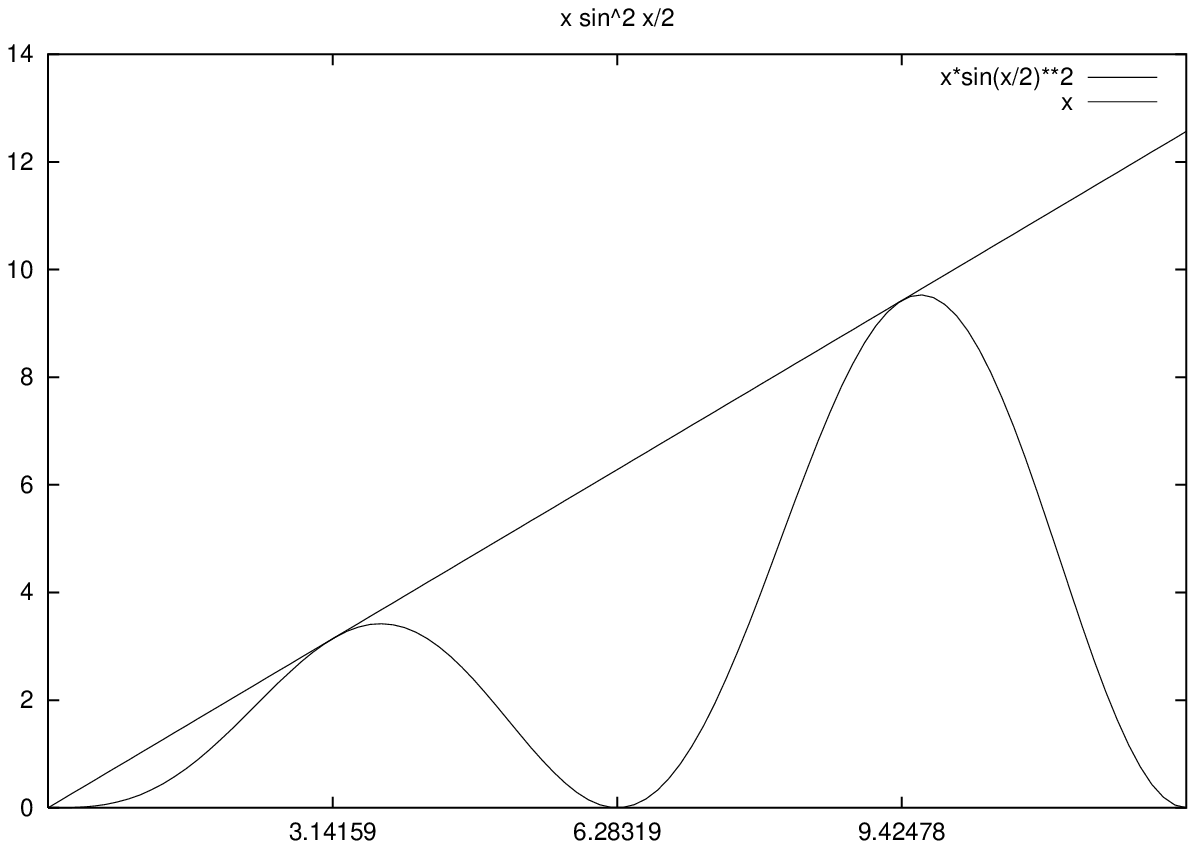}
\end{picture}
\begin{picture}(7,5)(0,0)
\put(0,4){$\displaystyle  \Delta_{31}\sin\Delta_{31}$}
\put(6,-0.2){$\displaystyle  \Delta_{31}$}
  \includegraphics[width=6cm,height=4cm,clip]{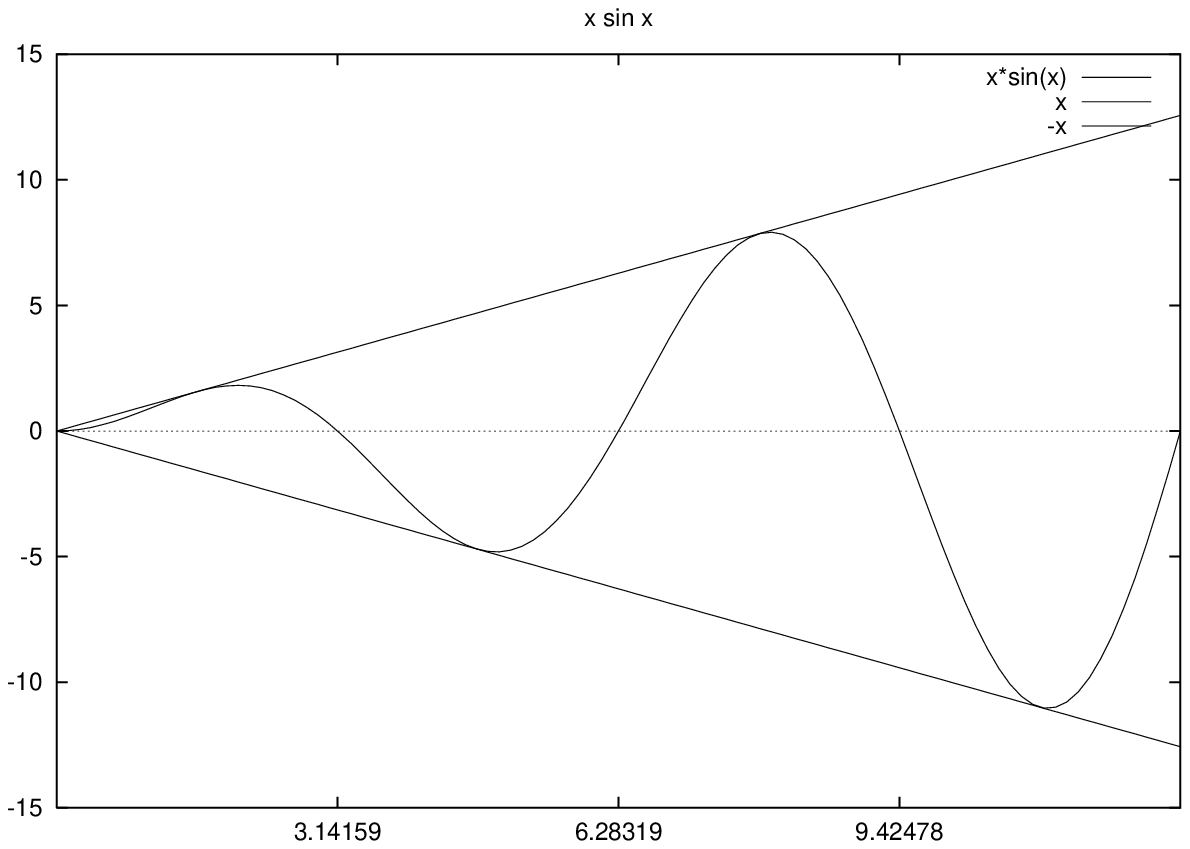}
\end{picture}
\begin{picture}(7,5)(0,0)
\put(0,4){$\displaystyle  \Delta_{31}^2$}
\put(6,-0.2){$\displaystyle  \Delta_{31}$}
  \includegraphics[width=6cm,height=4cm,clip]{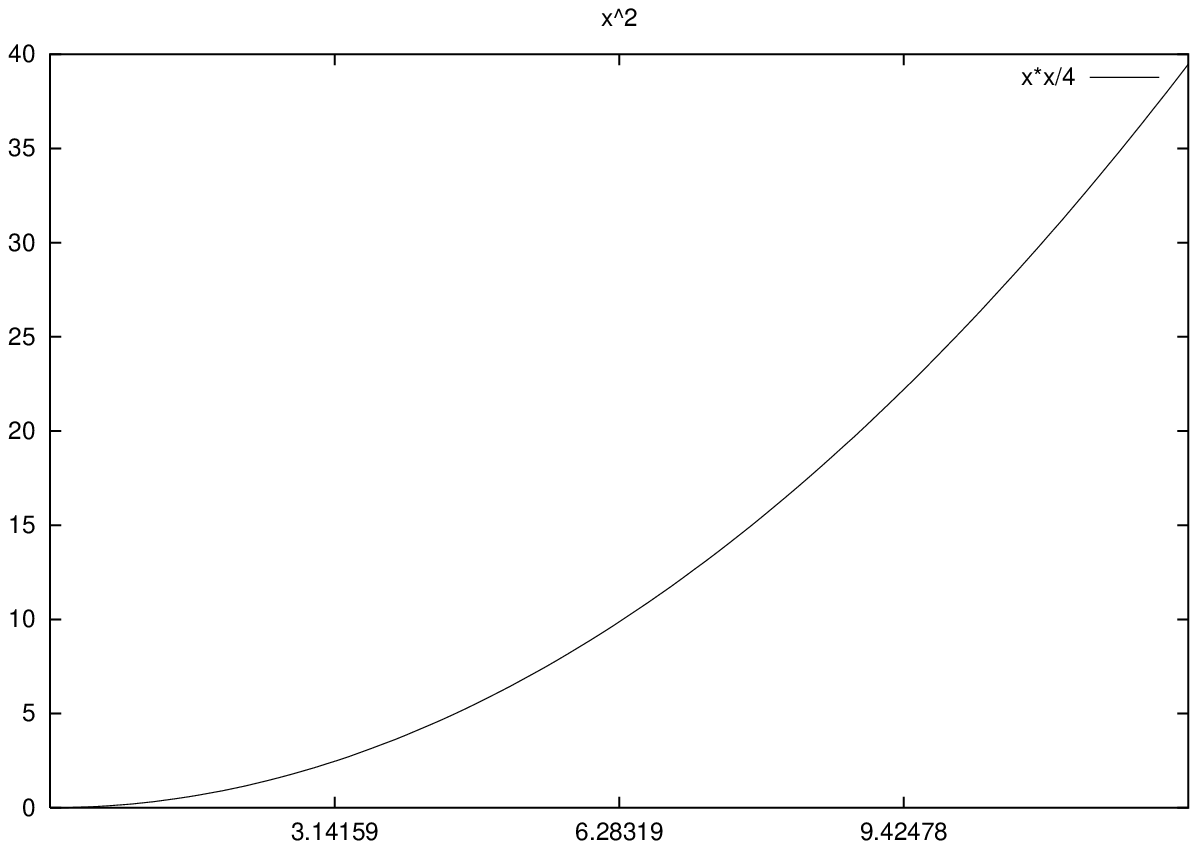}
\end{picture}
\caption{Base functions for transition probability eq.(\ref{transitionP}).}
\label{4schemes}
\end{figure}

\begin{figure}[h]
\unitlength 1cm
\begin{picture}(12,8)(0,0)
\put(12,0){$\displaystyle  \sin \delta$}
 \includegraphics[width=12cm,clip]{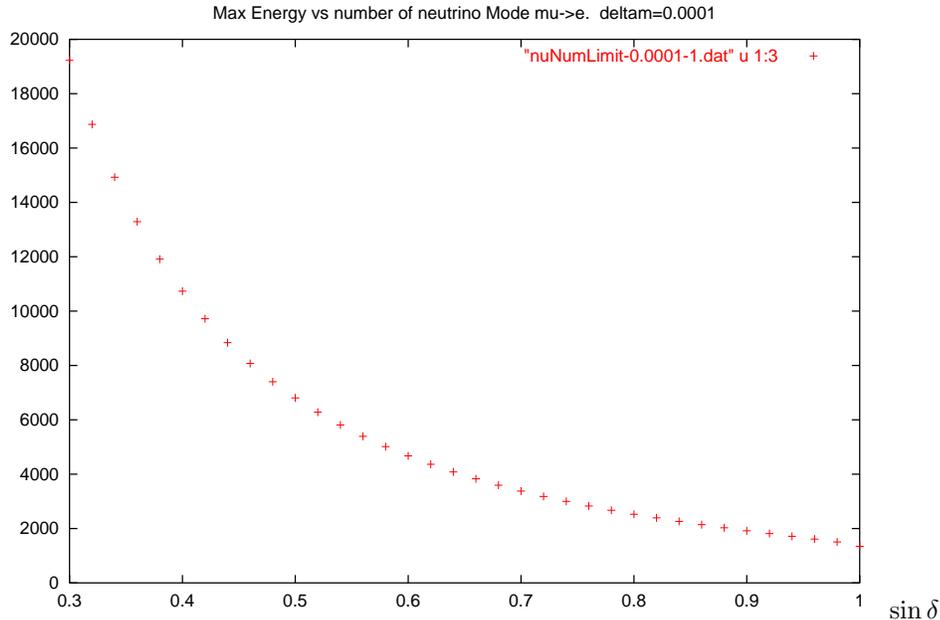}
\end{picture}
\caption{The necessary numbers of neutrinos for observing CP violation
as a function of $\sin\delta$ at 99\% level. The parameters here
are $\delta m^2_{21}=1\times 10^{-4}, \delta m^2_{31}=3\times 10^{-3},
,\sin 2\theta_{12}=0.8, \sin 2\theta_{23}=1 $ and $\sin 2\theta_{13}=0.2$.}
\label{need}
\end{figure}

In fig.\ref{need}, it is shown how many neutrinos and antineutrinos
in detection are necessary to see CP violation when all the mixing
parameters except CP violating phase are determined precisely
as indicated in the caption.
According to Konaka\cite{Konaka}, for SK size detector (20kt detector)
we will have 1000 $\mu$ neutrinos in detection per year with
the first stage of Japan Hadron Facility(JHF).
Then if $\sin\delta$ is large, we  can detect CP violation
with several years run without before neutrino factory runs.

There is another fruit using the current setting.
Though these 4 coefficients seem to be independent,
The following relation between coefficients
\begin{eqnarray}
4 A D = B^2+C^2
\label{Urelation}
\end{eqnarray}
must be satisfied if there are only 3-generation
neutrinos.
\footnote{Exactly speaking, this relation holds
up to $\frac{\delta m^2_{21}}{\delta m_{31}^2}$.
If we know the value of $\frac{\delta m^2_{21}}{\delta m_{31}^2}$,
then
\begin{eqnarray}
4 A\ (\tilde D -\frac{\tilde B}{2}) = {\tilde B}^2+{\tilde C}^2,
\nonumber
\end{eqnarray}
where
\begin{eqnarray}
B&=&\tilde B \times \frac{\delta m^2_{21}}{\delta m_{31}^2}\nonumber\\
C&=&\tilde C \times \frac{\delta m^2_{21}}{\delta m_{31}^2}\nonumber\\
D&=&\tilde D \times (\frac{\delta m^2_{21}}{\delta m_{31}^2})^2\nonumber
\end{eqnarray}
}
In other words we may check the unitarity of lepton sector.

\section{High energy? Low energy?}

As noted in the previous section,
what we observe are not angles of a certain parameterization
but values of certain combination of couplings.
We should verify which combination of the couplings
are determined well in an oscillation experiment and how 
we extract CP phase essentially. By this consideration
we can understand what kind of uncertainty in experiment
affects the uncertainty of determining angles.
Here we will consider how CP violation is observed in an oscillation
experiment as an example of this idea.

\begin{figure}[h]
\unitlength 1cm
\begin{picture}(12,7)(0,0)
\put(0,1){
\centerline{
  \includegraphics[width=7cm,height=5cm,clip]{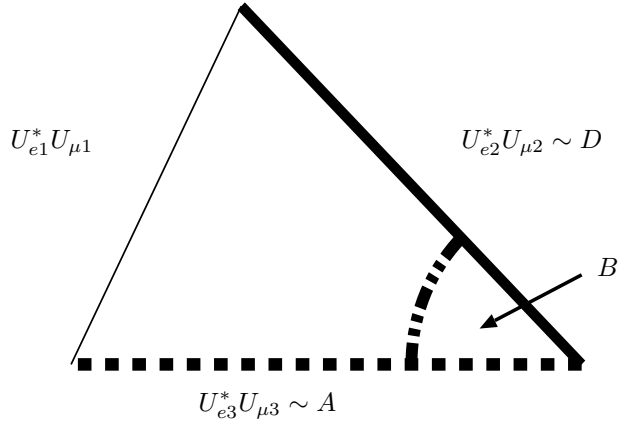}
}
}
\put(2,4){$\displaystyle  U_{e1}^*U_{\mu 1}$}
\put(8,4){$\displaystyle  U_{e2}^*U_{\mu 2} \sim D$}
\put(4.5,0.5){$\displaystyle  U_{e3}^*U_{\mu 3} \sim A$}
\put(9.8,2.3){$\displaystyle  B$}
\end{picture}
\caption{Unitarity triangle for lepton sector. This area shows
the strength of CP violation $C$. The length of the bottom line
is determined by $A$, that of right line corresponds to $D$
and the angle between them is obtained by $B$.}
\label{Utriangle}
\end{figure}

To observe CP violation means to measure the area of
the unitarity triangle of the lepton sector.
To measure the area there are two ways:
1) Direct measurement.
2) First determining the triangle then calculating it.
The first way is strong against other uncertainties,
those of other parameters, experiments and so on,
since whether there is CP violation is determined
by the fact that the area is not 0.

Indeed we have two ways of the determination.
The observables $A-D$ corresponds the elements of the unitarity triangle
indicated in fig.\ref{Utriangle}. $C$ is the area itself.
Thus to determine $C$ corresponds to direct measurement.
On the other hand, the length of the bottom line
is determined by $A$, that of right line corresponds to $D$
and the angle between them is obtained by $B$. With these three
parameters the triangle are fixed completely and the
area is calculated according to the unitarity relation,
eq.(\ref{Urelation}).

Which of ``direct measurement'' or ``unitarity''
 determines the CP violation essentially?

In higher energy region, the transition probability takes the form,
\begin{eqnarray}
P ( \nu_\mu \rightarrow \nu_e ) = (A+B+D) \Delta_{31}^2 + C \Delta_{31}^3
+ \cdots.
\nonumber
\end{eqnarray}
Thus an experiment in higher energy
becomes sensitive to only the combination\footnote{%
Of course, the matter effect distinguishes
these observables, though the separation is weaker than
the case considered in section 2.} of $A+B+D$ 
and the direct CP measure, $C$, becomes less
determined. It means CP violation is measured by the unitarity relation,
eq.(\ref{Urelation}) and hence the determination of CP violation
is easily influenced by uncertainty in the experiment.

On the contrary, in lower energy region $C$ is a good observable
and hence we can tell whether CP violation is there rather strictly.

With this consideration how important the 3-generation property
is. The best energy range for CP violation is in the range given
in eq.(\ref{bestEnergy}).

\section{ Discussion}

To see CP violation, we have to see 3-generation of neutrinos
simultaneously and hence the energy range,
$\delta m^2_{21} L \leq E \leq  \delta m^2_{31} L.$,
is found to be preferable for it.
More to say to avoid the matter effect the shorter
baseline length is better.
Indeed an experimental setup, 
\begin{eqnarray}
 E \sim  {\rm O(100) MeV}\ 
{\rm and}\   L \sim {\rm O(100)km}, 
\label{desire}
\end{eqnarray}
is very feasible and
much richer information on $\nu'$s can be obtained.
In this region we can see not only transition but also
 full oscillation and the observables are the combination of
couplings $A-D$ in eq.(\ref{transitionP}).

Large part of parameter space which will be probed
by neutrino factory can be surveyed by the conventional beam
with the setup (\ref{desire}).
Why don't you consider such a possibility seriously?

\section*{Acknowledgments}
The author thanks a. Konaka and M. Koike for useful discussions.
This
research was supported in part by a Grant-in-Aid for Scientific
Research of the Ministry of Education, Science and Culture,
\#12047221, \#12740157.


\end{document}